\documentclass[pre,twocolumn,showpacs]{revtex4}

\usepackage{amsmath,amssymb}
\usepackage{graphicx}
\usepackage{pslatex}

\begin{document}

\def\be{\begin{equation}}
\def\ee{\end{equation}}
\def\bfi{\begin{figure}}
\def\efi{\end{figure}}
\def\bea{\begin{eqnarray}}
\def\eea{\end{eqnarray}}

\title{Measure synchronization in a two-species bosonic Josephson junction}

\author{ Jing Tian$^{1,2}$}
\author{Haibo Qiu$^{1,2,3}$}\email[Email: ]{phyqiu@gmail.com}
\author{Guanfang Wang $^{3}$}
\author{ Yong Chen$^2$}
\author{Li-bin Fu$^{2,3}$}
\email[Email: ]{lbfu@iapcm.ac.cn}
\address{1. College of Science, Xi'an University of Posts and Telecommunications,
$710121$, Xi'an , China}
\address{2.Institute of Theoretical Physics, Lanzhou University,
$730000$, Lanzhou , China}
\address{3.Institute of
Applied Physics and Computational Mathematics, 100088, Beijing,
China}

\date{\today}

\begin{abstract}

Measure synchronization (MS) in a two-species bosonic Josephson
junction (BJJ) is studied based on semi-classical theory. Six
different scenarios for MS, including two in the Josephson
oscillation regime (0 phase mode) and four in the self-trapping
regime ($\pi$ phase mode) , have been clearly shown. Systematic
investigations of the common features behind these different
scenarios have been performed. We show that the average energies of
the two species merge at the MS transition point. The scaling of the
power law near the MS transition has been verified, and the critical
exponent is 1/2 for all of the different scenarios for MS. We also
illustrate MS in a three-dimensional phase space; from this
illustration, more detailed information on the dynamical process can
be obtained. Particularly, by analyzing the Poincar\'e sections with
changing interspecies interactions, we find that the two-species BJJ
exhibits separatrix crossing behavior at MS transition point, and
such behavior depicts the general mechanism behind the different
scenarios for the MS transitions. The new critical behavior found in
a two-species BJJ is expected to be found in real systems of atomic
Bose gases.

\end{abstract}


\maketitle

\section{\label{sec1}INTRODUCTION}

Coupled dynamical systems can show a magnificent collective behavior
called synchronization \cite{Review}, a concept first being
experimentally shown by Huygens with two marine pendulum clocks in
1665. In recent decades, many nontrivial features have been revealed
\cite{Diss}. In most such studies, coupled dissipative oscillators
are employed whereas research on coupled non-dissipative Hamiltonian
systems is still at a primitive stage because of complications
originating from Liouville's theorem \cite{MS06,MS97,JJ1,JJ2}. In
the latter system, a new kind of collective phenomenon called
measure synchronization (MS) was found. As demonstrated by Hampton
and Zanette \cite{MS99}, two coupled Hamiltonian systems experience
a dynamical phase transition from a state in which the two
Hamiltonian systems visit different phase-space domains to a state
in which the two Hamiltonian systems cover an identical phase-space
domain as the coupling strength increases. Such phenomena were later
investigated in coupled Duffing-, $\varphi_{4}$-, and
Frenkel-Kontorova-type Hamiltonian systems\cite{MS03,MS05,MS12}.

Experimentally, the superconducting Josephson junction (SJJ) is
perhaps the most widely studied class in the exploration of
synchronization; the superconducting Josephson junction can serve as
a prime example of coupled dynamical systems. With recent
experimental progress in Bose-Einstein condensates (BECs), a bosonic
Josephson junction (BJJ) can be created and controlled by confining
single-species BECs in a double well \cite{Exp}. In a pioneering
theoretical study \cite{Smerzi97}, Smerzi {\sl et al.} mapped a
single-species BJJ to a classical pendulum system. Therefore, it is
natural to expect that a two-species BJJ, which consists of a
two-species BEC provides a model system to study coupled dynamical
systems.

The single-species BJJ is of great significance in its own right.
The generalized Josephson equations describing the BJJ differ from
the ones used for the superconducting Josephson junction by the
presence of a nonlinear interaction term \cite{Smerzi97}. Because of
this term, a single-species BJJ can exhibit a counter-intuitive
phenomenon called macroscopic quantum self-trapping (MQST). In a
detailed analysis of this novel phenomenon \cite{Smerzi99}, the
Josephson oscillation (JO) regime and MQST regime can be seen in a
phase-plane portrait. Additionally, through an increase in the
nonlinear interaction term, the dynamical phase transition from JO
to MQST will occur because of the separatrix crossing behavior in
the phase space \cite{Zib10,Hen12,Kra09,Chuchem10}. This dynamical
phase transition behavior has been studied extensively,  both
theoretically and
experimentally\cite{Zib10,Hen12,Kra09,Chuchem10,Sakmann09,Polkovnikov10,Juli10,Fu06,Exp1}.

Theoretical analysis has been extended to a two-species BJJ
\cite{Ashhab02,xu08,
mazzarella09,satija09,diaz09,sun09,naddeo10,molmer12,Bur11,Ng05,mazzarella10,mazzarella11,Chat12,lu11}.
A system of equations for coupled pendula can be derived for the
temporal evolution of the relative population and relative phase of
each species. Many interesting tunneling effects have been found,
including the symmetry restoring phase \cite{satija09},
mixed-Rabi-Josephson oscillation \cite{mazzarella11}, counterflow
superfluidity \cite{mazzarella10}, and so on. We have studied
collective modes in a two-species BJJ \cite{Qiu}. In addition to
phase synchronization, we determined that measure synchronization
can also occur. The transitions between different modes can be found
by varying the interspecies interaction strength.

In this paper, we perform a systematic investigation on the measure
synchronization found in such systems. Six different scenarios for
MS are clearly determined. We identified that MS is a continuous
phase transition, that the scaling law for the MS transitions was
numerically verified, and that the critical exponent is $1/2$.
Particularly, separatrix crossing has been revealed to be the
dynamical mechanism behind the different scenarios for MS by
Poincar\'e section analysis. Because experimental progress has been
made in the production of two-species BECs with tunable intra- and
interspecies interactions \cite{Fesh,Fesh2}, we expect that a two
species BJJ can be realized and that the MS can thus be
experimentally investigated in the near future.

This paper is organized as follows. A brief description of a
two-species BJJ model is given in Section \ref{sec2}. In Sections
\ref{sec31}, different scenarios of MS are introduced. Section
\ref{sec6} presents a detailed analysis of different MS scenarios.
Conclusions are given in Section \ref{sec7}.

\section{The model\label{sec2}}

 A two-species bosonic Josephson junction (BJJ) can be experimentally realized by
trapping a binary mixture of BECs in a symmetric double well
potential. By assuming the interaction among the atoms is
sufficiently weak, with the well-known two-mode approximation
\cite{Smerzi97,Smerzi99,Legg01}, the Hamiltonian in the second
quantization reads:
\begin{eqnarray}
\hat{H} &=&\frac{u_{a}}{2N_{a}}[(\hat{a}_{L}^{\dag
}\hat{a}_{L})^{2}+(\hat{a}_{R}^{\dag
}\hat{a}_{R})^{2}]+\frac{u_{b}}{2N_{b}}[(\hat{b}_{L}^{\dag
}\hat{b}_{L})^{2}+(\hat{b}_{R}^{\dag
}\hat{b}_{R})^{2}]  \nonumber\\
&&-\frac{v_{a}}{2}(\hat{a}_{L}^{\dag }\hat{a}_{R}+\hat{a}_{R}^{\dag }\hat{a}%
_{L})-\frac{v_{b}}{2}(\hat{b}_{L}^{\dag }\hat{b}_{R}+\hat{b}_{R}^{\dag }\hat{b}_{L})   \nonumber\\
&&+\frac{u_{ab}}{\sqrt{N_{a}N_{b}}}(\hat{a}_{L}^{\dag
}\hat{a}_{L}\hat{b}_{L}^{\dag }\hat{b}_{L}+\hat{a}_{R}^{\dag
}\hat{a}_{R}\hat{b}_{R}^{\dag }\hat{b}_{R}),
\end{eqnarray}
where $\hat{a}_{L(R)}^{\dag }$ $(\hat{a}_{L(R)})$ and $\hat{b}_{L(R)}^{\dag }$%
$(\hat{b}_{L(R)})$ are the creation (annihilation) operators for the
localized modes in the left $(L)$ or right $(R)$ well of different
species($a$ or $b$) respectively. $N_{a}$ and $N_{b}$ stand for the
particle numbers of species $a$ and $b$.
$u_\sigma=(4\pi\hbar{\rm{a}}_\sigma N_\sigma /m_\sigma )\int
{\left|{\varphi_\sigma}\right|}^4 dr$, $u_{ab}=
2\pi\hbar{\rm{a}}_{ab}\sqrt{N_a N_{b}} (\frac{1}{{m_a}}+
\frac{1}{{m_{b}}})\int {\left|{\varphi _a} \right|}^2 \left|{\varphi
_{b}} \right|^2 dr $ denote the effective interaction of atomic
collision between the same kind of species and between the different species, respectively, with  $%
\sigma =a, b$ as the indication of the species, the interactions can
be either repulsive or attractive, depending on the sign of $u$.
Both $u_a$, $u_b$ and $u_{ab}$ can be tuned by Feshbach technique,
as demonstrated by experiments in a mixture of $^{87}$Rb and
$^{85}$Rb \cite{Fesh}.  $v_\sigma = \int {[(\hbar ^2 /2m_\sigma
)\nabla \varphi _L \nabla \varphi _R + V(r)\varphi _L \varphi _R ]}
dr$ is the effective Rabi frequency describing the coupling between
two wells.

Under the semi-classical limit \cite{Smerzi97,Smerzi99,Legg01},
dynamics of
the system can be described by a classical Hamiltonian $H=\langle \Psi _{GP}|%
\hat{H}|\Psi _{GP}\rangle /N$, in which $|\Psi _{GP}\rangle =\frac{1}{\sqrt{%
N_{a}}}(\alpha _{L}\hat{a}_{L}^{\dag }+\alpha _{R}\hat{a}_{R}^{\dag
})^{N_{a}}|0,0\rangle \bigotimes \frac{1}{\sqrt{N_{b}}}(\beta _{L}%
\hat{b}_{L}^{\dag }+\beta _{R}\hat{b}_{R}^{\dag
})^{N_{b}}|0,0\rangle $ is the
collective state of the N-particle system with $N=N_{a}+N_{b}$. Here, $%
\alpha _{j}=|\alpha _{j}|e^{i\theta _{aj}}$ and $\beta _{j}=|\beta
_{j}|e^{i\theta _{bj}}$ $(j=L$ or $R)$ are four $c$ numbers which
correspond to the probability amplitudes of the two different
species of atoms in the two wells. And the conservation of particle
numbers of each species requires: $|\alpha _{L}|^{2}+|\alpha
_{R}|^{2}=1,\ $ $|\beta _{L}|^{2}+|\beta _{R}|^{2}=1$.

   By introducing the relative population difference: $S_{a}=(|\alpha
_{L}|^{2}-|\alpha _{R}|^{2}),\ $ $S_{b}=(|\beta _{L}|^{2}-|\beta
_{R}|^{2})$, and the relative phases difference $\theta _{\sigma
}=\theta _{\sigma L}-\theta _{\sigma R}$ . We obtain the mean-field
Hamiltonian \cite{Ashhab02},
\begin{eqnarray}
\label{hamil1} H_{tot}&=&H_{a}+H_{b}+H_{I},
\end{eqnarray}
it is composed of Hamiltonian $H_{\sigma}$  ($\sigma$=$a$,$b$)
\begin{eqnarray}
\label{hamil2}
H_{\sigma}&=&\frac{u_{\sigma}}{2}S_{\sigma}^{2}-v_{\sigma}\sqrt{1-S_{\sigma}^{2}}\cos\theta_{\sigma},
\end{eqnarray}
and the coupling term
\begin{eqnarray}
\label{hamil3} H_{I}&=&u_{ab}S_{a}S_{b}.
\end{eqnarray}

$H_{\sigma}$ is well-known as the mean field Hamiltonian for a
single-species BJJ \cite{Smerzi97,Smerzi99}; $H_{I}$ is the coupling
term. Thus, a two-species BJJ is similar to two coupled
single-species BJJs. It is clear that the coupling occurs because of
the presence of the interspecies interaction $u_{ab}$.

The equations of motion can be derived by computing:
 $\dot{\theta_{\sigma}}=\frac{\partial H}{\partial
S_{\sigma}}$, $\dot{S_{\sigma}}=-\frac{\partial H}{\partial
\theta_{\sigma}}$, we obtain:

\begin{equation}
\label{fdspect1}
\dot{\theta_{a}}=u_{a}S_{a}+\frac{v_{a}S_{a}}{\sqrt{1-S_{a}^{2}}}\cos\theta_{a}+u
_{a b }S_{b}
\end{equation}
\begin{equation}
\dot{S_{a}}=-v_{a}\sqrt{1-S_{a}^{2}}\sin\theta_{a}
\end{equation}
\begin{equation}
\label{fdspect2}
\dot{\theta_{b}}=u_{b}S_{b}+\frac{v_{b}S_{b}}{\sqrt{1-S_{b}^{2}}}\cos\theta_{b}+u
_{ab }S_{a}
\end{equation}
\begin{equation}
\dot{S_{b}}=-v_{b}\sqrt{1-S_{b}^{2}}\sin\theta_{b}.
\end{equation}

The tunneling dynamics of a two-species BJJ can be described with
Eqs. (5)-(8). Here, the standard fourth-order Runge-Kutta method is
used to obtain a numerical solution. Because we are interested in
showing the effects of coupling on the dynamics of each species, the
collective motions are presented by projecting the state of the full
system onto the individual phase spaces, i.e., we study the
trajectories ($S_{a}(t)$ ,$\theta _{a}(t)$) in the phase plane
$(S_{a},\theta_{a})$ and the trajectories ($S_{b}(t),\theta
_{b}(t)$) in the phase plane $(S_{b},\theta_{b})$.

The tunneling dynamics for a single-species BJJ have been
extensively studied
\cite{Smerzi97,Smerzi99,Legg01,Sakmann09,Polkovnikov10,Juli10,Chuchem10,Fu06,Exp1},
and studies of $H_{\sigma}$ based on the semi-classical theory have
shown that there are two distinct dynamic regimes in phase space
\cite{Smerzi97,Smerzi99,Fu06}: the Josephson oscillation regime, and
the self-trapping regime with a strong nonlinearity ($u/v
> 1$). For simplicity, $0$-phase will be used to stand for the Josephson oscillation, in which $\theta_{\sigma}$ oscillates around
$\theta_{\sigma}=0$. And $\pi$-phase stand for the self-trapping, in
which $\theta_{\sigma}$ oscillates around $\theta_{\sigma}=\pi$. To
show the coupled dynamical behavior of $H_{\sigma}$, we will then
categorize the initial configurations of a two-species BJJ into two
broad categories:

  (i)  $0$-phase mode

  (ii) $\pi$-phase mode

\begin{figure}
\begin{center}
\includegraphics[width=0.8\columnwidth,angle=0, clip=true]{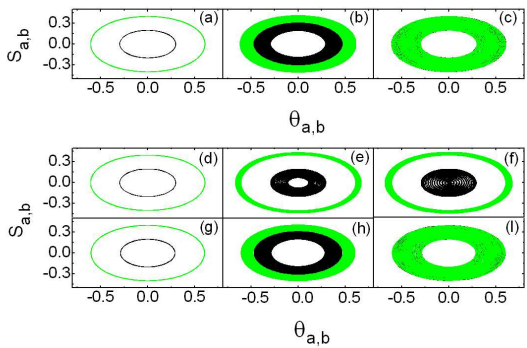}
\caption[]{(color online)  Phase-space domains of the two species in
the $0$ phase mode. The two species are represented by green and
black.  Initial configuration
($S_{a}$,$S_{b}$,$\theta_{a}$,$\theta_{b}$) set to be (0.2, 0.4, 0,
0).  (a) $u_{ab}=0$. (b) $u_{ab}$=0.0086. (c) $u_{ab}$=0.009: MS is
achieved. (d) $u_{ab}=0$. (e) $u_{ab}$=-0.01. (f) $u_{ab}$=-0.0325.
(g) $u_{ab}$=-0.0625. (h) $u_{ab}$=-0.0738. (i) $u_{ab}$=-0.08: MS
is achieved.
 } \label{fig1}
\end{center}
\end{figure}

\section{\label{sec31}MS in $0$- and $\pi$-phase mode}

\subsection{\label{sec3} $0$-phase mode}

First, we present measure synchronization in the $0$-phase mode,
that is, the mode in which $\theta_{\sigma}$ oscillates around
$\theta_{\sigma}=0$. The initial conditions
($S_{a}$,$\theta_{a}$,$S_{b}$,$\theta_{b}$) are (0.2, 0.0, 0.4,
0.0), $u_{a}=u_{b}=1.2$, and $v_{a}=v_{b}=1$.

\subsubsection{\label{sec3b}With repulsive interactions ($u_{ab}>0$)}

Figures~\ref{fig1} $(a)$-$(c)$ show the evolution process for MS
with repulsive interspecies interactions ($u_{ab}>0$). By increasing
the coupling strength $u_{ab}$, we draw orbits on the
($S_{\sigma},\theta_{\sigma}$), $(\sigma=a,b)$ phase plane of the
two subsystems. For $u_{ab}=0$, as shown in Fig.~\ref{fig1}(a),
these initial conditions correspond to two different quasiperiodic
orbits, which cover closed curves in green and black. For
$u_{ab}>0$, the two closed curves are replaced with two smooth
quasi-periodic trajectories wandering in two distinctive phase-space
domains, which are ring shaped. As $u_{ab}$ increases, the two
phase-space domains first evolve such that the external border of
the inner domain approaches the internal border of the outer domain,
and the two approach each other until $u_{ab}=0.0086$, at which
point the two approaching boundaries are almost in contact
[Fig.~\ref{fig1}(b)]. Then, a sudden change occurs as $u_{ab}$
increases further, as shown in Fig.~\ref{fig1}(c). The two formally
well-separated phase-space domains merge and cover the phase-space
domains with identical invariant measure \cite{MS99}. This dynamical
phase transition of the two phase-space domains marks the transition
to measure synchronization. The evolution process described above is
identical to that described previously \cite{MS99,MS03,MS05}, which
we will call typical MS hereafter.

\subsubsection{\label{sec3c}With attractive interactions ($u_{ab}<0$)}

Figures~\ref{fig1}$(d)$-$(i)$ show the evolution process with
increasing strength of attractive interspecies interactions. As a
starting point, in Fig.~\ref{fig1}(d), we plot the orbits for each
species at zero coupling. For $u_{ab}<0$, as $u_{ab}$ decreases, we
see that this evolution process is quite different from the typical
MS. The two phase-space domains first evolve in the opposite
direction; the internal border of the inner ring approaches the
center of the phase-space domain, whereas the external border of the
outer ring expands [Fig.~\ref{fig1}(e)]. For $u_{ab}=-0.0325$, the
internal border of the inner ring finally reaches the center of the
phase space [Fig.~\ref{fig1}(f)]. Then, as $u_{ab}$ decreases, these
two rings gradually thinning until $u_{ab}=-0.0625$; at this point,
the two rings again become two curves[Fig.~\ref{fig1}(g)], which
appear similar to $u_{ab}=0$ [Fig.~\ref{fig1}(d)]. Additionally, as
$u_{ab}$ continues to decrease [Figs.~\ref{fig1} $(g)$-$(i)$], the
evolution process becomes identical to that for the typical MS
process [as described in Figs.~\ref{fig1} $(a)$-$(c)$].

\subsection{Localized $\pi$-phase mode\label{sec4}}

Here, we present measure synchronization in $\pi$-phase mode, in
which $\theta_{\sigma}$ oscillates around $\theta_{\sigma}=\pi$. The
initial conditions ($S_{a}$,$\theta_{a}$,$S_{b}$,$\theta_{b}$) are
(0.2, $\pi$, 0.4, $\pi$), $u_{a}=u_{b}=1.2$, and $v_{a}=v_{b}=1$.

\subsubsection{\label{sec4b}With repulsive interaction ($u_{ab}>0$)}

\begin{figure}
\begin{center}
\includegraphics[width=0.8\columnwidth,angle=0, clip=true]{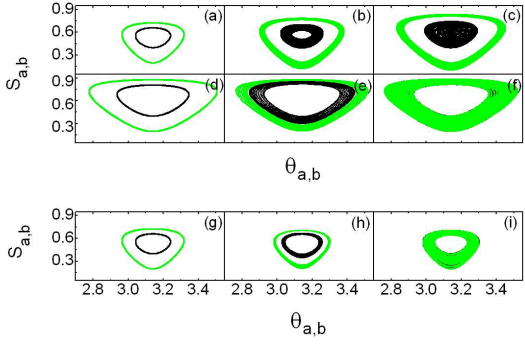}
\caption[]{(color online)  Phase-space domains of the two species in
the localized $\pi$ phase mode. The two species are represented by
green and black. Initial configuration
($S_{a}$,$S_{b}$,$\theta_{a}$,$\theta_{b}$) set to be (0.2, 0.4,
$\pi$, $\pi$). (a) $u_{ab}=0$. (b) $u_{ab}$=0.03. (c)
$u_{ab}$=0.0737. (d) $u_{ab}$=0.1498. (e) $u_{ab}$=0.1621. (f)
$u_{ab}$=0.1622: MS is achieved for repulsive $u_{ab}$. (g)
$u_{ab}=0$. (h) $u_{ab}$=-0.01. (i) $u_{ab}$=-0.0123 MS is achieved
for attractive $u_{ab}$.
 } \label{fig3}
\end{center}
\end{figure}

Figures~\ref{fig3} $(a)$-$(f)$ show the evolution process for MS
with repulsive interspecies interactions ($u_{ab}>0$). For
$u_{ab}=0$, as shown in Fig.~\ref{fig3}(a), these initial conditions
correspond to two different quasiperiodic orbits, which cover closed
curves in green and black, and the two curves have an
inverted-triangle shape. By increasing the coupling strength, we can
see that the two inverted triangles broaden in such a way that the
two embedded phase-space domains evolve in opposite directions
[Fig.~\ref{fig3}(b)], and the internal border of the original inner
inverted triangle approaches the center of the phase space. At
$u_{ab}=0.0737$, the inner phase-space domain reaches the center
[Fig.~\ref{fig3}(c)]. Subsequently, the two phase-space domains
approach one another until they make contact before the MS
transition at $u_{ab}=0.1621$, and in this evolution process, there
is also be a moment at which the phase-space domains become closed
curves again [Fig.~\ref{fig3}(d)]. This scenario is similar to the
scenario for attractive interactions in the $0$-phase mode. The most
obvious difference is that the phase-space domains no longer have
conserved boundaries; as the coupling strength increases, the area
expands.

\subsubsection{\label{sec4c}With attractive interaction ($u_{ab}<0$)}

Figures~\ref{fig3} $(g)$-$(i)$ show the scenario with increasing
strength of attractive interspecies interactions. This scenario is
very similar to the scenario for typical MS. We find that as the
coupling strength increases, the two phase-space domains approach
each other until MS occurs at $u_{c}=-0.0123$.

\subsection{Nonlocalized $\pi$-phase mode\label{sec5}}

\begin{figure}
\begin{center}
\includegraphics[width=0.8\columnwidth,angle=0, clip=true]{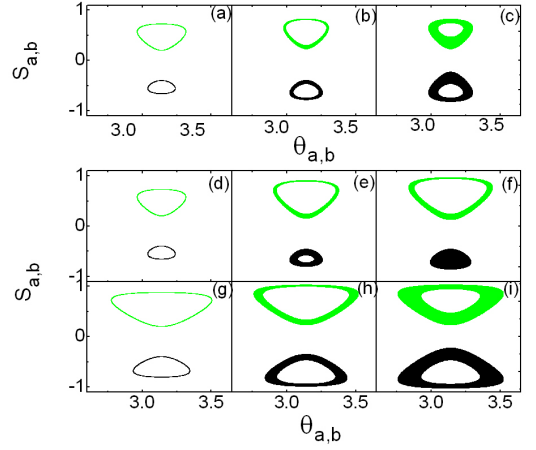}
\caption[]{(color online) Phase-space domains of the two species in
the nonlocalized $\pi$ phase mode.  The two species are represented
by green and black. With the initial conditions
($S_{a}$,$\theta_{a}$,$S_{b}$,$\theta_{b}$) taken to be (0.2, $\pi$,
-0.4, $\pi$) in $\pi$-phase mode. (a)$u_{ab}=0$. (b) $u_{ab}$=0.01.
(c) $u_{ab}$=0.0123: nonlocal MS is achieved for repulsive $u_{ab}$.
(d) $u_{ab}=0$. (e) $u_{ab}$=-0.03. (f) $u_{ab}$=-0.0737. (g)
$u_{ab}=-0.1498$. (h) $u_{ab}$=-0.1621. (i) $u_{ab}$=-0.1622:
nonlocal MS is achieved for attractive $u_{ab}$.
 } \label{fig5}
\end{center}
\end{figure}

In the $\pi$-phase mode, a new type of coherent evolution process is
found, as shown in Fig.~\ref{fig5}. For the initial conditions
($S_{a}$,$\theta_{a}$,$S_{b}$,$\theta_{b}$) of (0.2, $\pi$, -0.4,
$\pi$). At $u_{ab}=0$, compared with the localized $\pi$-phase mode,
these initial conditions also correspond to two closed curves but
with one curve on top of the other [Fig.~\ref{fig5}(a)].

As the strength of repulsive interspecies interactions increase, the
phase-space domains of the two species become more comparable in
area until $u_{ab}$ reaches a critical value ($u_{ab}=0.0123$)
[Fig.~\ref{fig5}(b)]; then, a sudden change occurs, as shown in
Fig.~\ref{fig5}(c), and the two phase-space domains have the same
area. However, in contrast to Fig.~\ref{fig3}(i), the phase-space
domain of each species lies symmetrically on both sides of the line
$S=0$ [Fig.~\ref{fig5}(c)].

As the strength of attractive interspecies interactions increases,
we find another scenario for the transition behavior that ends in a
similar state [Fig.~\ref{fig5}(i)]. Interestingly, we note that this
scenario has many features in common with the scenario shown in
Figs.~\ref{fig3} $(a)$-$(g)$. One major difference is the structure
of the phase-space domains: one structure goes from top to bottom,
whereas the other structure is embedded.

Comparing Fig.~\ref{fig3} and Fig.~\ref{fig5}, we can see that each
phase diagram corresponds to $u_{ab}$ values with the same
magnitudes but with the opposite signs. This result can be
understood by analyzing Equation (2)-(4); if we set $s_{a}$ and
$s_{b}$ to have opposite signs and let $u_{ab}$ also have a value
with opposite sign, the coupling term $H_{I}$ does not change and
neither $H_{a}$ or $H_{b}$. The two different initial conditions
with opposite signs for the interspecies interactions correspond to
the same Hamiltonian and, consequently, have the same dynamic
evolution.

\section{Analyses\label{sec6}}

Below, we will explore the nature of the MS found for a two species
BJJ in detail.

\subsection{The Energy Characteristics}

For the two groups of MS scenarios that have been found, the $0$-
and $\pi$-phase modes, we analysis the energy function for each
species: ${\rm{E}}_{{\rm{a,b}}}$, and observe how the energy
function changes with interspecies interactions. Here
 \begin{equation}
\label{fdspect3} {\rm{E}}_{{\rm{a,b}}}=
 \frac{u_{a,b}}{2}S^{2}_{a,b}-v_{a,b}\sqrt{1-S^{2}_{a,b}}\cos\theta_{a,b}.
\end{equation}

\begin{figure}
\begin{center}
\includegraphics[width=0.9\columnwidth,
angle=0, clip=true]{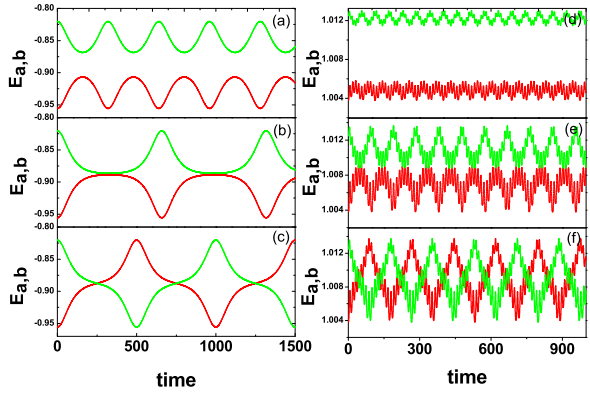} \caption[]{(color online) Evolution of
energy functions for the two species. The left column is for the
$0$-phase mode and right column for the $\pi$-phase mode. Before MS
is achieved $(u_{ab}<u_{c})$, the two species have different energy
variations; after MS is achieved $(u_{ab}\geq u_{c})$, the two
energy variations would be the same. The de-mixing to mixing feature
of MS transitions can be clearly seen. } \label{fig6}
\end{center}
\end{figure}

In Fig.~\ref{fig6}, we plot the energy function for each species
with different interspecies interaction strengths below $u_{c}$ for
the initial configuration in the $0$-phase mode and $\pi$-phase mode
for repulsive interspecies interaction. Before measure
synchronization, ${\rm{E}}_{{\rm{a}}}$ and $ {\rm{E}}_{{\rm{b}}}$ do
not overlap at all. As $u_{ab}$ continues to increase, the
difference between the lower boundary of the initially higher energy
species and the upper boundary of the initially lower energy species
becomes smaller and smaller. When $u_{ab}$ reaches the transition
point $u_{c}$, ${\rm{E}}_{{\rm{a,b}}}$ suddenly has the same range
of energy variations. This evolution process is shown in
Figs.~\ref{fig6}$(a)$-$(c)$ for the $0$-phase mode and in
Fig.~\ref{fig6}$(e)$-$(g)$ for the $\pi$-phase mode. The de-mixing
to mixing feature of the MS transitions can be clearly seen in these
Figures.

To describe MS in the context of our physical model, the average
energy of a single-species BJJ is defined to be:
\begin{equation}
\label{fdspect4} <{\rm{E}}_{{\rm{a,b}}}
>  = \frac{{\rm{1}}}{{\rm{T}}}\int_{\rm{0}}^{\rm{T}}
{{\rm{E}}_{{\rm{a,b}}}
 {\rm{dt}}}
\end{equation}

\begin{figure}
\begin{center}
\includegraphics[width=0.8\columnwidth,
angle=0, clip=true]{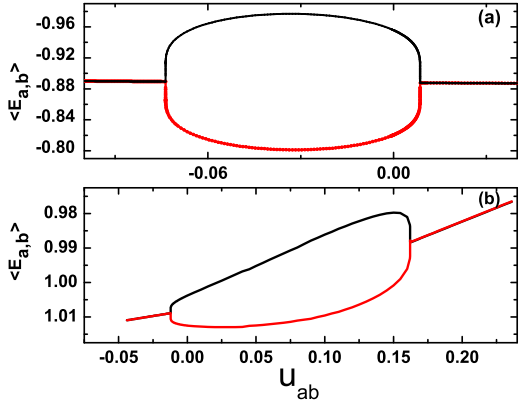} \caption[]{(color online)  The average
energies of the two species in the $0$- and $\pi$-phase mode. The
two subsystems would have equal averaged energy once MS is achieved.
(a). In the $0$-phase mode. (b). In the $\pi$-phase mode. }
\label{fig7}
\end{center}
\end{figure}

In Fig.~\ref{fig7}(a), we show the averaged energy $<E_{a}>$ and
$<E_{b}>$ as a function of interspecies interactions $u_{ab}$ in the
$0$-phase mode. It is clear that there are sharp transitions at
$u_{ab}=u_{c}=0.0086$ and $u_{ab}=u_{c}=-0.0738$ for the repulsive
and attractive interactions, respectively. Below $u_{c}$
$(|u_{ab}|<|u_{c})|$, there is a finite difference between $E_{a}$
and $E_{b}$, whereas above $u_{c}$ $(|u_{ab}|>|u_{c})|$, both
species have identical average energies. Fig.~\ref {fig7} (b) shows
the plot for the $\pi$-phase mode. The correspondence of the MS
transition with the sudden merging of the average energies is also
clearly shown. In addition, we find that the average energies change
in the $\pi$ phase mode, even in the measure-synchronized state,
whereas in the $0$-phase mode, the average energies remain fixed.

\subsection{The critical behavior}

The critical behavior of MS has been studied previously. In the
seminal work \cite{MS99}, Hampton and Zanette introduced an order
parameter to study the critical logarithmic singularity; however,
they did not find the scaling law and the critical exponent because
of the order parameter, which is an averaged quantity, that was
chosen for the calculation. In Ref.
 \cite{MS03}, through the computation of the interaction energy and the phase dynamics
of the oscillators, the scaling law behind MS in coupled
$\varphi_{4}$ systems was extensively discussed, and different
scaling laws were numerically verified before and after MS. The
critical exponents are $1/3$ and $1/2$.

\begin{figure}
\begin{center}
\includegraphics[width=0.9\columnwidth,
angle=0, clip=true]{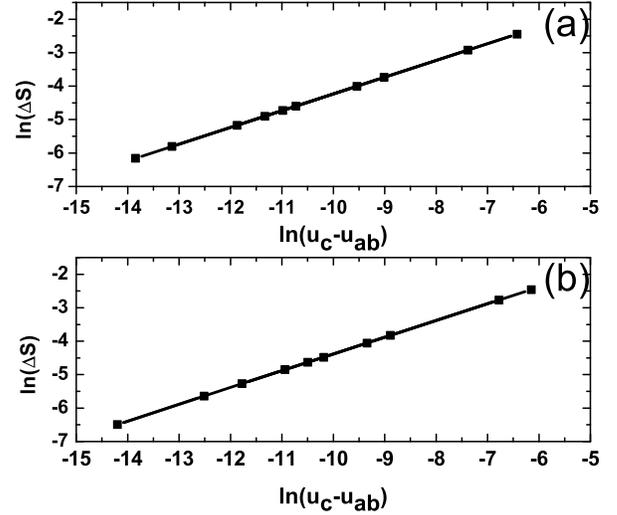} \caption[]{(color online) Scaling
relation of  $\triangle S\propto(u_{ab}-u_{c})^{\frac{1}{2}}$ for
the $0$-phase mode with repulsive interaction (a), and $\pi$-phase
mode with repulsive interaction (b).} \label{fig61}
\end{center}
\end{figure}

Here, we studied the critical behavior of MS in a two-species BJJ.
We confirmed that there are scaling laws in this system. As the two
phase-space domains approach each other, we noticed that the two
phase-domain boundaries are getting close to contact, and there is a
scaling relation behind this process. By computing the distance
between ($\triangle S$) the two approaching boundaries on the
$S_{a,b}$ axes, we find the scaling relation between $\triangle S$
and $(u_{ab}-u_{c})$; this relation is $\triangle
S\propto(u_{ab}-u_{c})^{\frac{1}{2}}$, and the critical exponent is
$1/2$. Fig.~\ref{fig61} (a) shows the scaling relation for the $0$
phase mode with repulsive interactions and Fig.~\ref{fig61} (b)
shows the scaling relation for the $\pi$ phase mode with repulsive
interactions. For the other scenarios, we have verified that the
critical exponents are all identical and are $1/2$.

\subsection{Three-dimensional description}

Previous work on MS only studied the two-dimensional projected
phase-space domains of the coupled-Hamiltonian system. However, this
projection could not be a complete description of the dynamical
behavior because the dynamics of the two-coupled Hamiltonian
actually take place in a four-dimensional phase space
$(S_{a},\theta_{a},S_{b},\theta_{b})$. In the absence of
dissipation, energy constrains the motion of the system to a
three-dimensional energy hypersurface of the four-dimensional phase
space. To gain the most insight into MS, we can use a
three-dimensional description of MS.

\begin{figure}
\begin{center}

\includegraphics[width=0.15\columnwidth]{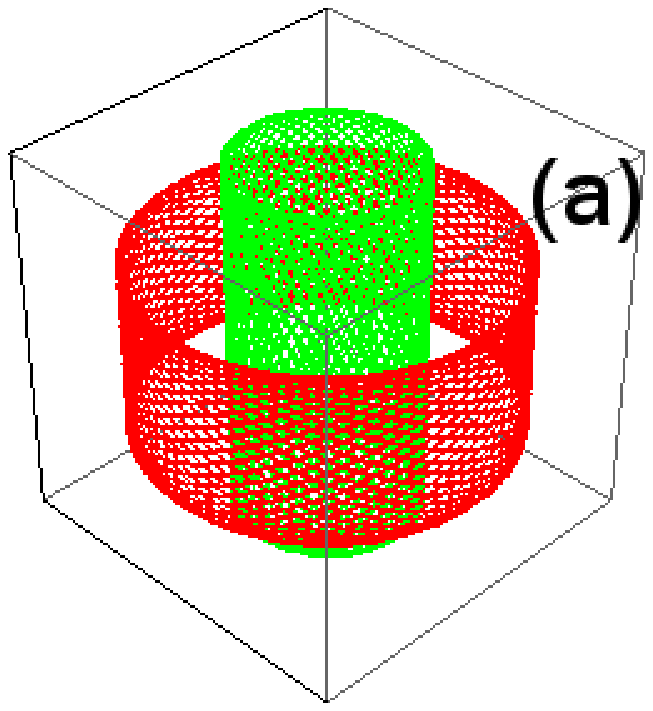}
\includegraphics[width=0.15\columnwidth]{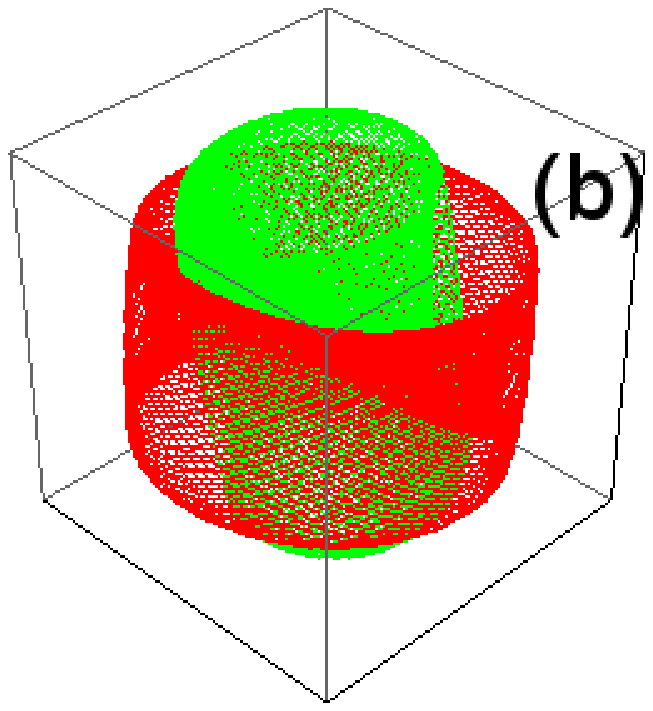}
\includegraphics[width=0.15\columnwidth]{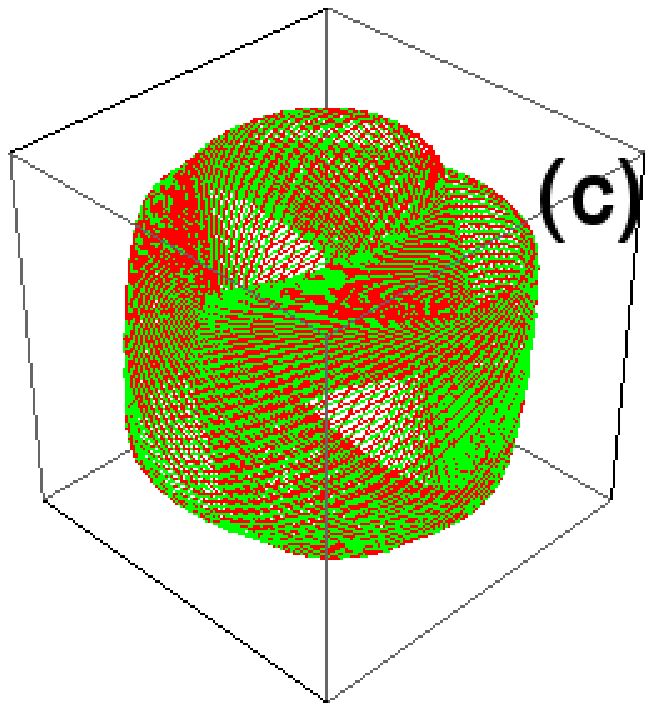}

\includegraphics[width=0.15\columnwidth]{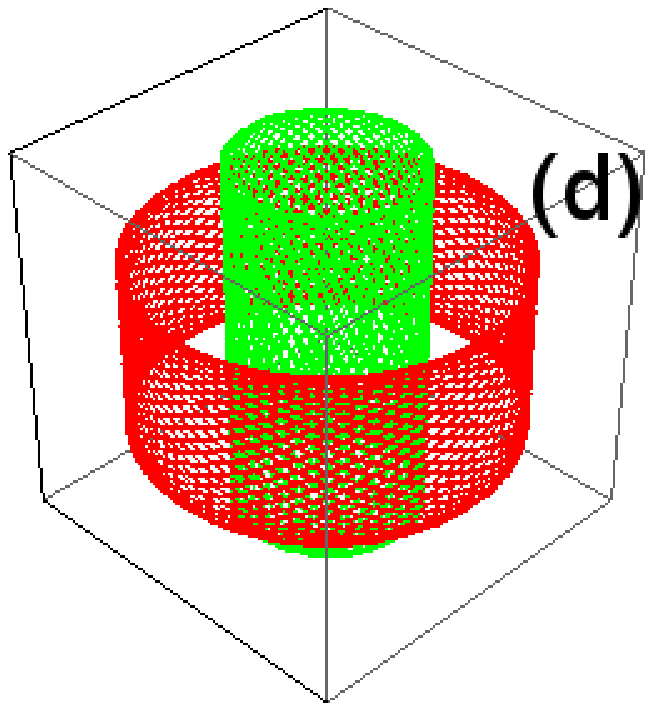}
\includegraphics[width=0.15\columnwidth]{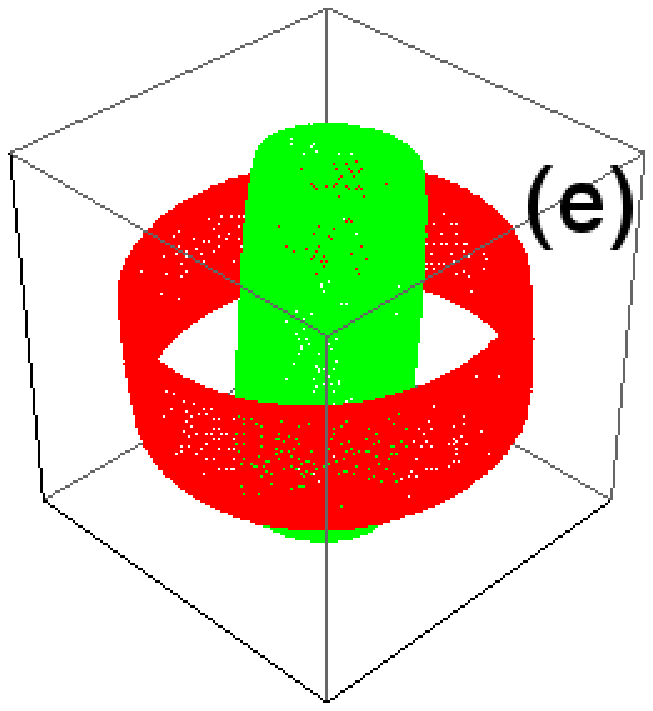}
\includegraphics[width=0.15\columnwidth]{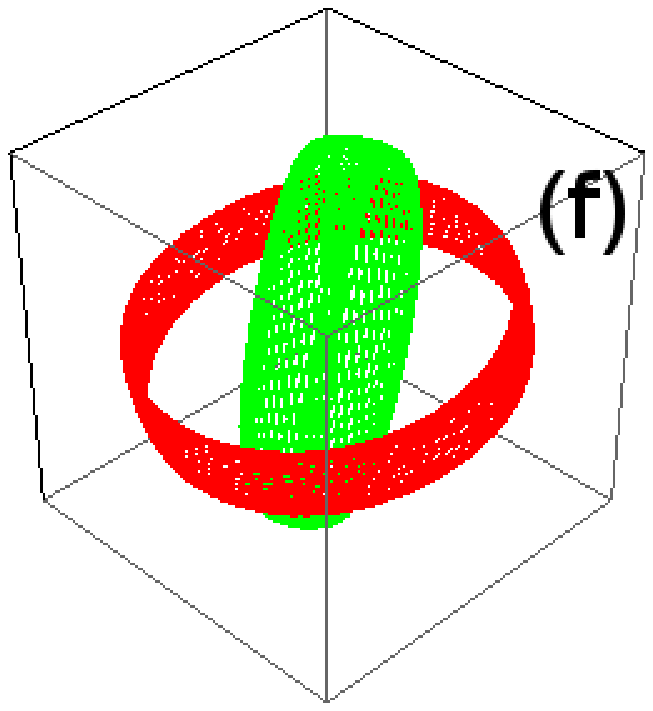}

\includegraphics[width=0.15\columnwidth]{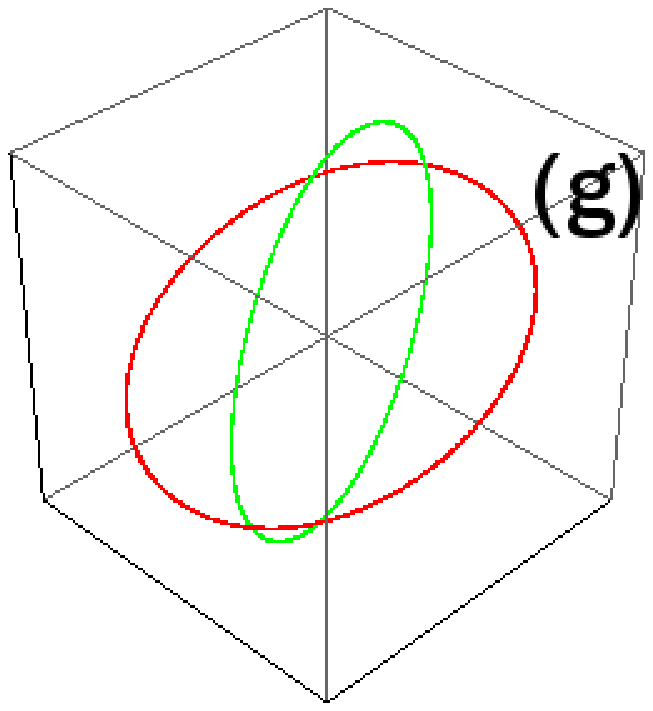}
\includegraphics[width=0.15\columnwidth]{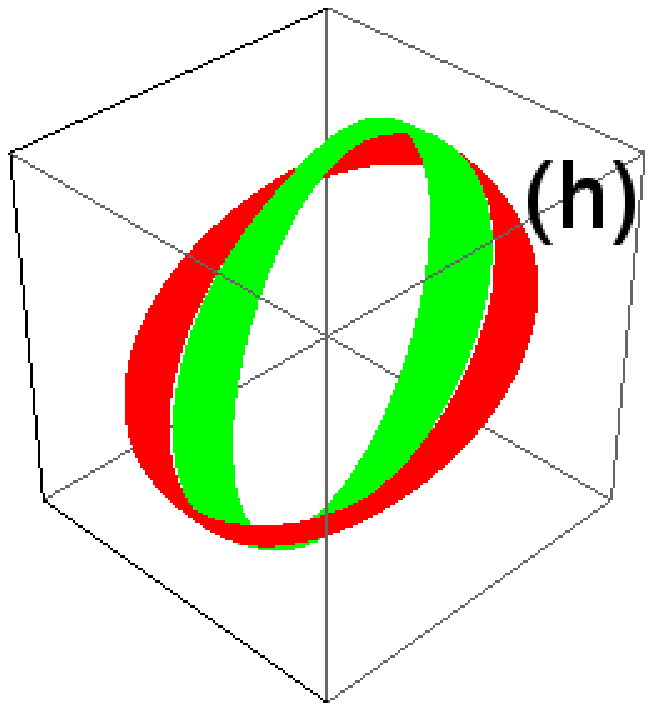}
\includegraphics[width=0.15\columnwidth]{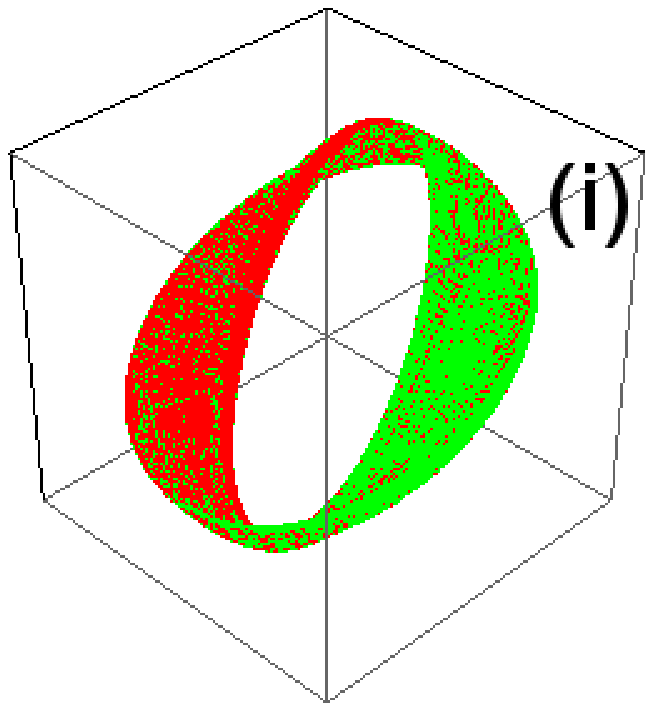}

\caption{(Color online)  A three-dimensional view of MS for a
two-species BJJ in the $0$ phase mode. Two different colors
represent two different choices of the three-dimensional axes; the
green one is drawn on axes ($S_{a}$,$\theta_{a}$,$\theta_{b}$) and
the red one is drawn on axes ($S_{b}$,$\theta_{b}$,$\theta_{a}$).
(c) and (i) show measure synchronized states.} \label{fig2}
\end{center}
\end{figure}

By taking the initial configuration in the $0$-phase mode as an
example, we provide a three dimensional description of measure
synchronization in Fig.~\ref{fig2}. First, we choose two different
sets of coordinate axes:
$(S_{\sigma},\theta_{\sigma},S_{\bar{\sigma}})$ and
$(S_{\bar{\sigma}},\theta_{\bar{\sigma}},S_{\sigma})$, with $\sigma
=a,b$. The corresponding initial conditions and the coupling
$u_{ab}$ are identical to those in Fig.~\ref{fig1}.

With repulsive interspecies interactions ($u_{ab}>0$), the evolution
process of the three-dimensional phase space is shown in
Figs.~\ref{fig2}$(a)$-$(c)$; these figures show side views of the
two manifolds in the three-dimensional phase space representation
and top views corresponding to the figures shown in
Figs.~\ref{fig1}(a),(b),(c). In Fig.~\ref{fig2}(a), for $u_{ab}=0$,
there are two well-separated manifolds, with one manifold around the
other. For $u_{ab}$ =0.0086, as shown in Fig.~\ref{fig2} (b), the
two manifolds are close to each other but are still separated well
from one another. However, with $u_{ab}>0.0086$, as shown in
Fig.~\ref{fig2} (c) $(u_{ab}=0.009)$, we see that the two manifolds
completely overlap; this overlap indicates the measure-synchronized
states in the three-dimensional phase space representation.

With attractive interspecies interactions ($u_{ab}<0$), the
evolution process of the three-dimensional phase space is shown in
Figs.~\ref{fig2}$(d)$-$(i)$. The process shown in
Figs.~\ref{fig2}$(d)$-$(i)$ is not as direct as in case of the
repulsive interspecies interactions, because initially the inner
phase-space volume shrinks in size (Figs.~\ref{fig2}$(d)$-$(g)$),
then, this volume expands continuously until it achieves the
measure-synchronized states (Fig.~\ref{fig1}(i)). There, we can see
some behaviors that are not apparent on the two-dimensional map;
e.g., although Fig.~\ref{fig1}(g) and Fig.~\ref{fig1}(d) appear to
be exactly identical, they are actually very different, as shown in
the three-dimensional representation: Fig.~\ref{fig2}(d) shows
quasi-periodic states, whereas Fig.~\ref{fig2}(g) shows periodic
states. The volume of the synchronized state also apparently
changed; however, in the 2-D projection, we cannot see many of these
changes.

  To summarize,  a three-dimensional view of MS is given. It is observed that as MS is
attained, the two energy manifolds in the phase space
($(S_{\sigma},\theta_{\sigma},S_{\bar{\sigma}})$ completely overlap.
This result provides a more intuitive picture of MS compare with the
2-D projection, and some features that we do not see in the
two-dimensional phase space are presented. These features include
the difference between the quasiperiodic state and periodic state
[Fig.~\ref{fig1}(d) and Fig.~\ref{fig1}(g)], and the changing volume
of the manifolds as $u_{ab}$ increases can be seen clearly. These
results help us understand the measure invariance of the
two-dimensional phase-space domains after MS is achieved because the
phase-space domains can be seen to be the projection of the two
energy manifolds on a two-dimensional phase plane.

\subsection{Poincar\'e Section Analysis}

Measure synchronization is a dynamical phase transition phenomenon
in coupled Hamiltonian systems, so naturally, we ask how this
phenomenon occurs? We find that the answer can be revealed through
the analysis of the Poincar\'e maps of the system.

\begin{figure}
\begin{center}
\includegraphics[width=1.0\columnwidth,
angle=0, clip=true]{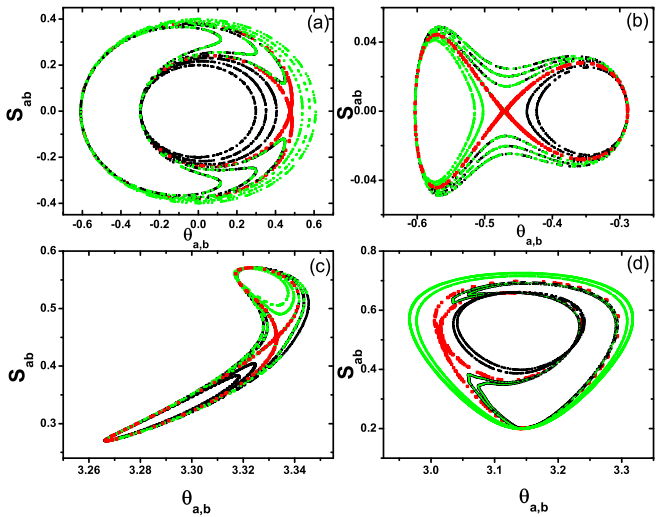} \caption[]{(color online)
Poincar\'e section analysis. The upper figures and lower figures
show the Poincar\'e section analysis for scenarios in
Fig.~\ref{fig1} and in Fig.~\ref{fig3}, respectively. The red curves
with X-point geometry mark the MS transitions, which correspond to
separatrices.} \label{fig8}
\end{center}
\end{figure}

The procedure for our analysis can be demonstrated for the example
of the repulsive interactions in the $0$-phase mode
[Fig.~\ref{fig8}(a)]. First, we solve the canonical
equations(5)-(8); then, we take the section slice of
($S_{a}$,$\theta_{a}$) at each time for which $\theta_{b}=0.0$ and
$\dot{\theta}_{b}>0$, the section slice taken in this procedure is
marked with black dots. Simultaneously, we also take the section
slice of ($S_{b}$,$\theta_{b}$) at each time for which
$\theta_{a}=0.0$ and $\dot{\theta}_{a}>0$, and this type of section
slice is marked with green dots. In Fig.~\ref{fig8}(a), different
curves with the same color are drawn for different values of
$u_{ab}$ that we chose. For the black dotted curves, with
$u_{ab}=0.001$, the corresponding Poincar\'e section is the
innermost, closed, ring-shaped curve. As the coupling intensity
increases, this ring-shaped curve expands until $u_{ab}$ reaches
$u_{c}$ ($u_{c}=0.008621$); at $u_{c}$, the section slice
corresponds to the separatrix, which is marked with red dots. For
$u_{ab}>u_{c}$, the section slice is shaped like a crescent moon and
shrinks in size as $u_{ab}$ increases further. The green dotted
curves are drawn for the same chosen set of $u_{ab}$ values; the
outermost curve corresponds to the Poincar\'e section for
$u_{ab}=0.001$. Conversely, we observe that this ring-shaped
Poincar\'e section shrinks in size before $u_{ab}$ reaches $u_{c}$,
and at $u_{c}$, the separatrix is also shown; beyond $u_{c}$, the
ring-shaped curve also assumes a crescent moon shape and shrinks in
size as $u_{ab}$ increases further. We note that the separatrix
marks the transition from the localized to the shared of the phase
space. After the measure synchronized of the two coupled Hamiltonian
systems for $u_{ab}>u_{c}$, the green-dotted and black-dotted
trajectories with the same coupling intensities merge completely.

Similarly, we perform Poincar\'e section analysis for the other
scenarios. Fig.~\ref{fig8}(b) shows the result for the $0$-phase
mode with attractive interspecies interactions, and
Figs.~\ref{fig8}(c) and (d) show the results of Poincar\'e section
analysis for the $\pi$-phase mode with attractive interspecies
interactions and repulsive interspecies interactions, respectively.
We can see that in all cases separatrices mark the onset of measure
synchronization.

In summary, through Poincar\'e section analysis, we have shown that
a two-species BJJ exhibits separatrix crossing behavior at the
critical interspecies interaction of the MS transition. Therefore,
we identified the separatrix crossing to be the underlying dynamical
mechanism of MS.

\section{Conclusion\label{sec7}}

To summarize, measure synchronization in a two-species BJJ has been
systematically studied. Six different scenarios of MS, including two
in the $0$-phase mode, four in the localized and nonlocalized
$\pi$-phase mode, have been characterized and some common features
behind have been revealed. We have found that the MS transition
correspond to the sudden mergence of average energies of the two
species. The power law scaling behind the MS transition has been
verified, which is the same for the different scenarios, and the
critical exponent is $1/2$. Furthermore, we have given a
three-dimensional view of MS which provides a more intuitive picture
of MS. And some features which we will not see in the two
dimensional phase space are revealed. In particular, by using the
Poincar\'e section analysis, it has been clearly shown that a
two-species BJJ exhibits separatrix crossing behavior at $u_{c}$ .
We conclude that separatrix crossing is the general mechanism behind
the different scenarios of MS transitions found in the two species
BJJ.

\begin{acknowledgments}
We thank Qiang Gu for critical reading of the manuscript. This work
was supported by the National Natural Science Foundation of China
(No.11104217, No.11205121), the Science Plan Foundation office of
the Education Department of Shaanxi Province (No.11JK0555), the
Youth Foundation of XUPT under Grant No. 0001295, No. 0001287.
\end{acknowledgments}

\end{document}